\documentclass[11pt,twoside]{article}

\usepackage{asp2004}
\usepackage{epsf}
\usepackage{psfig}
\usepackage{lscape}

\def\gtsima{$\; \buildrel > \over \sim \;$}    
\def\gtrsim{\lower.5ex\hbox{\gtsima}}           
\def\ltsima{$\; \buildrel < \over \sim \;$}    
\def\lesssim{\lower.5ex\hbox{\ltsima}}           
\def\pcc{{\rm ~cm}^{-3}}

\def\Pwnm{P_{\rm WNM}}
\def\sun{{\rm sun}}

\begin{document}
\title{Small Ionized and Neutral Structures: A Theoretical Review}   
\author{Steven R. Spangler$^1$ and Enrique V\'azquez-Semadeni$^2$}   
\affil{ $^1$Department of Physics and
Astronomy, University of Iowa, $^2$Centro de Radioastronom\'\i a y
Astrof\'\i sica, Universidad Nacional Aut\'onoma de M\'exico}    

\begin{abstract} 
The workshop on Small Ionized and Neutral Structures in the Interstellar
Medium featured many  contributions on the theory of the objects which
are responsible for ``Tiny Scale Atomic Structures'' (TSAS) and
``Extreme Scattering Events'' (ESE).  The main demand on theory is
accounting for objects that have the high densities and small sizes
apparently required by the observations, but also persist over a
sufficiently long time to be observable. One extensively-discussed
mechanism is compressions by transonic turbulence in the warm
interstellar medium, 
followed by thermal instabilities leading to an even more compressed
state.  In addressing the requirements for overpressured but persistent
objects, workshop participants also discussed fundamental topics in the
physics of the interstellar medium, such as the timescale for
evaporation of cool dense clouds, the relevance of
thermodynamically-defined phases of the ISM, the effect of magnetic
fields, statistical effects, and the length and time
scales introduced by interstellar processes. 
\end{abstract}

\section{Introduction}   

The SINS meeting was dominated by an interesting and often disconnected
mix of observational reports on ``small scale'' structure in various
interstellar environments, and theoretical works attempting to explain
such observed structures. Interestingly, there was a wider variety in
the kinds and physical properties of the structures reported
observationally than in the theoretical approaches presented as
potential models.
This article includes a review of the theoretically-significant results
presented at the SINS workshop.  For reference, we first recall observational
results that pose the challenge for the theoretical studies.

\section{Classes of Compact Structure}

Small-scale structure was reported to be observed in all three of the
diffuse ionized, atomic and molecular components of the interstellar medium
(ISM), although the physical properties of the structures in each
component are as varied as the environments in which they are found (see
the observational review by Heiles \& Stinebring in this volume). Even
the very notion of ``small'' differs strongly from one field to
another. The features are primarily distinguished by their observational
signatures.
 
\subsection{Dense Regions of Ionized Gas} 

In the diffuse ionized gas, small-scale structures are inferred from radio 
scintillation effects,
such as large flux density variations of compact radio sources or
pronounced ``fringing'' in the dynamic spectra of pulsars.  A more
subtle observational indicator is a host of effects suggesting that
additional fluctuations in plasma density are needed above and beyond
the Kolmogorov spatial power spectrum which has become canonical.  Comments addressing the existence of such excess power were made in the presentations by Barney Rickett and Dan Stinebring.  

Deducing physical properties of these ionized enhancements is not so straightforward,
and more model-dependent than for the atomic and molecular components, 
but the results that are
available give densities $\sim 10$ cm$^{-3}$, sizes $\lesssim 1$ AU,
and filling factors $\sim 0.1$ (Rickett's talk). 

The so-called Extreme Scattering Events (ESEs) are, well, more extreme,
with estimated densities in the range $300 - 10^5$ cm$^{-3}$, and diameters of $0.06 - 0.38$ AU (Clegg, Fey, and Lazio 1998).  
The structures responsible for ESE could be more extreme versions of the structures responsible for the ubiquitous pulsar scintillation arcs, which also seem to involve discrete, identifiable clouds rather than a continuum of turbulent fluctuations (Stinebring's talk).

\subsection{``Tiny scale atomic structure'' (TSAS): High Density Neutral
Atomic Gas} \label{sec:tsas}

Most of the presentations dealt with observations of measurable changes
in the neutral hydrogen column density over small angular distances on
the sky. Interpreted at face value, these features would correspond to
small ($\sim $ 10-100 AU in size) regions of greatly enhanced
gas density (see Crystal Brogan's talk). Characteristic values of the gas
density would be $\sim 10^5$ cm$^{-3}$ as opposed to $\sim 0.5 \pcc$ for the
typical Warm Neutral Medium (WNM), or $\sim 50 \pcc$ for the Cold
Neutral Medium (CNM). Even with the assumption of low
neutral hydrogen temperatures of $\simeq 50$ K, these regions would then
exceed the typical pressure in the neutral medium by $\sim 3$ orders of
magnitude. The filling factor of TSAS is, however, very low, $\sim
0.001$. 

\subsection{Small-scale structure in molecular gas}

While attention at this workshop concentrated on the CNM, WNM, and DIG
(Diffuse Ionized Gas) components of the interstellar medium, small-scale molecular structures 
were also reported at the conference by Andreas Heithausen, with
typical densities $\sim 10^3$-$10^4 \pcc$, temperatures $\sim 10$-20 K,
sizes $\sim 100$-5000 AU, and masses $\sim 0.1$-1 $M_\sun$. Such masses 
are much smaller than their virial masses ($\sim 1 M_\sun$), implying 
that these structures cannot be gravitationally bound, and must be
formed or confined by some form of pressure (possibly ram
pressure). Edith Falgarone and Pierre Hily-Blant
further pointed to tiny molecular clouds in which 
the HCO$^+$ abundances are much higher than 
those expected in steady-state models and in which abundance ratios vary
wildly, suggesting a very dynamical state.  Edith interpreted these data in terms of turbulence-induced chemical reactions leading to HCO$^+$, resulting from localized heating caused by turbulent dissipation.  


\subsection{Theoretical challenges}

The main theoretical challenge in understanding these structures is
that, if they are true density enhancements, then they are
strongly overpressured with respect to the ambient ISM pressure, and,
naively, they should either not be there, or disperse quickly. Possible
explanations presented at the meeting included persistent turbulent
production of transient structures, geometrical effects, or statistical
observational projection effects. In \S \ref{sec:theor_interp} we summarize
these possibilities.

If the structures are really small-scale density and pressure
enhancements in the ISM, then the challenge is to
understand how one can generate such large excursions out of a medium with
typical densities between $\sim 0.1 \pcc$ (the Diffuse Ionized Gas or
DIG) and $\sim 1$ cm$^{-3}$ (the WNM). For the scintillation features in
ionized gas, the fact that this gas is ionized removes any possibility
of the pressure being reduced by a low gas temperature.

\section{Basic Theoretical Descriptions of the Interstellar Gas} 


Since the warm ISM is characterized by a transonic velocity dispersion,
theoretical explanations for the SINS as actual density enhancements
rely on transonic turbulence in a cooling, magnetized medium.  There are
basic questions about how one theoretically describes the gas
that comprises the interstellar medium, and different descriptions may be
necessary for the different phases.  Ellen Zweibel discussed the large
variety of spatial scales that arise in the gas due to viscosity,
resistivity, and collisions between ions and neutrals.  In addition,
there are still remaining basic physics issues about how one describes
the true viscosity in a magnetized, collisionless plasma.  The ESE
structures occur on (very roughly) the ion-neutral collisional scale in
the DIG, i.e. the scale that corresponds to the wavelength of an
Alfv\'{e}n wave with a frequency equal to the ion-neutral collision
frequency.  Perhaps such fundamental plasma scales define the sizes of
TSAS and ESE phenomena.

The study of turbulence in plasmas is also highly relevant to the small
structures in the interstellar medium, since the phenomena we observe
probably grow out of the general turbulence field.  If this is the case,
we need to fully understand some of the topics discussed by Alex
Lazarian, Stanislav Boldyrev and Joanne Mason, such as
the location and extent of the inertial subrange and dissipation range
of the turbulence, and the relationship between different turbulence
variables, like density and magnetic field.

The interstellar medium may provide a new arena for the study of plasma turbulence.  Much of plasma turbulence theory has depended on spacecraft
observations of the solar wind for data support.  While this
theory-observation link has proven fruitful, other turbulent media in
astrophysics might differ in important ways.  Solar wind turbulence at 1
AU is dynamically young, nearly collisionless, and
describable by a simple energy equation.  The interstellar medium is
different in each of these categories, so the lessons learned from
studies of the solar wind might not be applicable here.  Another
specific difference, which might illuminate fundamental differences, is
in the presence or absence of compressibility.  In solar wind
turbulence, the density fluctuations are small relative to those of
magnetic field and flow velocity.  
The ISM, on the other hand, is characterized by a range of densities
and temperatures spanning many orders of magnitude, and its cooling
properties render it highly compressible and even thermally unstable in
the range $8000~{\rm K} \gtrsim T \gtrsim 500~{\rm K}$.

\section{Theoretical interpretations} \label{sec:theor_interp}

\subsection{Numerical results}

A number of investigators presented numerical simulations of dynamical
compressions at moderate Mach numbers (${\cal M} \lesssim 2$) (Enrique
Vazquez-Semadeni, Patrick Hennebelle) or global transonic turbulence in
the WNM (Adriana Gazol), subject to thermal bistability 
\cite[e.g.,][]{FGH69}, showing that in 
principle a population of small-scale, high-density, high-pressure
objects can be transiently formed in such flows. The basic principle is
that the transonic compressions in the WNM nonlinearly trigger thermal
instability, causing a phase transition to a cold, dense phase
\cite{HP99}.  Since the total (thermal + ram) pressure is higher than
the thermal pressure of the WNM ($\Pwnm$) alone, the cold cloudlets end
up at pressures higher than $\Pwnm$ as well \cite{VRPGG06}. Moreover,
the compressed 
layers develop turbulence \cite[e.g.,][]{V94, WF98, KI02,
Hetal05, VRPGG06} and,
since they are much colder than the WNM ($T 
\lesssim 50$ K), this turbulence is strongly supersonic, so that the
flow there is highly compressible and the highest-density structures are
transient. 

In the simulations, pressures up to $P \sim 10^5$ K$\pcc$ and densities
up to $n 
\sim 10^3$ are readily reached (see Adriana's and Patrick's
contributions). It is noteworthy that structures of a given 
high density (e.g., $n \sim 300 \pcc$) can span roughly two orders of
magnitude in thermal pressure ($10^4$-$10^6$ K$\pcc$), indicating that
dense cloudlets can exist at both high and low temperatures.
Presumably, the warmer ones are those that have not had time to cool
down yet, and are probably in the process of doing so. For example, a
cloudlet at $n \sim 3000 \pcc$ and $P \sim 10^6$ K$\pcc$ is at a
temperature $T \sim 300$ K and cooling strongly. It is already much
cooler than the WNM, but not yet at the temperatures characteristic of
the CNM or the molecular gas. This suggests that indeed these structures
are transient, in particular the warmer ones.

In principle, then, dense, overpressured, small-scale structures (both 
cold and warm) can be generated in such flows. Patrick reported
densities up to $n \gtrsim 10^4 \pcc$ and pressures up to $P \lesssim
10^6$ K$\pcc$ in 2D simulations with a resolution of 10000 grid points
per dimension, down to scales of hundreds of AU. This appears very
close to explaining the physical conditions of \emph{individual}
structures reported by the atomic and molecular observations. However,
in order to determine whether the mechanism can account for the
\emph{entire population} of structures, detailed statistical comparisons
are necessary.  
Adriana reported 7\%
of the mass being at $n \ge 100 \pcc$, although Patrick warned that
this fraction seems to depend on resolution. In general, it is necessary to
measure this fraction, as well as the filling factors, as a function of
the threshold density used to define the clumps in the simulations in
order to compare in detail with the observations. This should allow a
quantitative determination of whether the mechanism of transient,
out-of-equilibrium density fluctuations by transonic turbulence in the
WNM can account for the \emph{population statistics} of these objects,
and not only the existence of a few such objects in isolation. Other means
of comparison based on the energy and density spectra as derived taking
into account projection effects were outlined by Alex Lazarian.

If comparisons prove to be statistically correct, then the mechanisms 
can be considered viable in principle, although much higher resolution
and the inclusion of the relevant microphysical processes will still be
needed for the simulations to reach the scales of the ionized structures
responsible for radio scintillation. On the other hand, if the
simulations do not produce these structures in sufficient numbers, then
it may be 
necessary to explain the remainder of the observed structures by other
mechanisms, such as those discussed below in \S\S
\ref{sec:geom} and \ref{sec:projec} 

 
\subsection{Implications for the Significance of Phases of the
Interstellar Medium} 

Much of the fundamental theoretical description of the interstellar
medium has been based on the concept of phases arising from thermodynamic equilibrium.  The identification of
the cold and warm phases of the interstellar medium with portions of the
equilibrium $P(\rho)$ relationship dates back nearly 40 years \cite{FGH69} and is
appealing because of its physical simplicity. However, there is
presently a controversy, expressed indirectly at this meeting by the
(sometimes loud!) discussions between Patrick Hennebelle and Enrique
V\'azquez-Semadeni, as to whether these
concepts are meaningful, or if the real interstellar medium is always so
far from thermodynamic equilibrium as to make those concepts
invalid. The physics of the actual ISM is probably a mixture of these concepts.  In
regions of strong turbulent mixing, the interfaces between the
equilibrium phases may be blurred, the flow being more similar to a
density continuum than to a two-phase medium. Conversely, in more
quiescent regions, thermal instability may be free to act and generate
two-phase structure. Indeed, the simulations show both kinds of
situations, with CNM cloudlets sometimes surviving for long times
relatively unperturbed, and some others rapidly dispersing away back
into the diffuse gas.

In regions where two-phase structure is present, studies of clump
evaporation are relevant, and were presented by Jonathan Slavin, Inoue
San and Nagashima San, with reports of typical evaporation timescales
$\sim 1$ Myr. However, since the small-scale structures that are the subject of this
conference are generally agreed to be strongly overpressured, it is
expected that their lifetimes are determined by a dynamical formation
and re-expansion process, with the characteristic timescale being their
turbulent crossing time, which is generally much shorter than the
evaporation time. In regions where two-phase structure is present, the
development of thermal instability may also generate moderate turbulence
(Inutsuka \& Koyama's poster).

One disappointment in the workshop was the lack of a more vigorous and
extensive discussion about the relationship (if any) between the atomic
and ionized entities. Barney Rickett and Mark Walker are to be complimented
as progressives in this respect. In coming to an eventual understanding
of these objects, it will be helpful to know if the radio propagation
effects are caused by the ionized outer envelopes of the clouds
responsible for TSAS, or if we are dealing with entirely distinct
structures that reside within the CNM or WNM on one hand, and the DIG on
the other.

\subsection{The Role of the Magnetic Field}

The proper primitive equations for study of the interstellar medium are
the equations of magnetohydrodynamics, not just hydrodynamics.  The
interstellar medium is permeated by a magnetic field with a magnitude of
3-5 $\mu$G, resulting in a plasma $\beta$ (defined as the square of the
ion-acoustic to Alfv\'{e}n speed ratio) of less than 1.  The magnetic
field pressure and tension play a dynamic role in the interstellar
medium, and cannot be ignored. 

Several presentations illustrated the role played by interstellar magnetism. Alex Lazarian discussed  some intrinsically magnetic effects, such as
the generation of density fluctuations by MHD waves even at scales
smaller than the viscous scales. Adriana Gazol presented results from
simulations of thermally bistable, magnetized turbulent flows indicating
that in this case a population of cold regions of low density (i.e., very
low thermal pressure) supported by magnetic pressure exists in the flow. 
The question then comes
down to how much variation in $n k_B T$ can be mechanically balanced by
$\frac{1}{c} \vec{J} \times \vec{B}$. The value of $\frac{B^2}{8 \pi}$ for $B=5 \mu G$ is $10^{-12}$ dynes/cm$^2$, corresponding to an $nT$ product of $7.2 \times 10^3$ K/cm$^3$.  Large variations in the magnetic energy density could therefore balance variations of several thousand (cgs units) in the dimensional quantity $nT$.  


It is widely believed  that, due to flux freezing, the magnetic field
should be enhanced together with the density fluctuations. However, both
observations \cite[e.g.,][ Crystal Brogan's talk at this meeting]{Crutcher99,HT03} and numerical
simulations of MHD turbulence \cite[e.g.,][]{PVP95, PN99, OSG01} show
otherwise.  The magnetic field strength appears to be in general uncorrelated with
the density, except perhaps at the highest densities.  In his talk,
Enrique V\'azquez-Semadeni recalled a proposed explanation for this
phenomenon \cite{PV03} based on the fact that the different modes of
\emph{nonlinear} MHD waves (so called ``simple'' waves) are
characterized by different scalings of the field strength with density,
so that in a turbulent medium the field strength at a given point does
not directly depend on the density, but on the (random) history of wave
passages through that point. It had been previously shown already
\cite{HP00} that compressions of a thermally bistable medium oblique to
the magnetic field up to a certain angle (that depends on the Mach
number of the compression) can induce a transition to the dense phase
without increasing the field, because the latter re-orients the motions
along itself, and the matter can slide freely along the field. Thus, the
lack of correlation is to be expected even in ideal MHD situations.


\subsection{Geometry effects?} \label{sec:geom}

In addition, and in some sense prior to this, we must be
certain that the observations really indicate relatively isotropic
regions of enhanced density, rather than highly anisotropic regions such
as sheets,  which are fortuitously
aligned with the line of sight. This possibility was first discussed by Carl Heiles in 1997. In this case, the implied volume densities 
and pressures would not be as extreme as the face-value implications of
the observations. Statistical analyses are again necessary to 
determine whether the chance alignment of the structures is sufficient
to account for the observed frequency of the phenomenon.

\subsection{TSAS and ESE as Statistical Fluctuations?} \label{sec:projec}

A. Deshpande proposed an explanation for TSAS and ESE that contrasts greatly with all of the other presentations, which attribute these phenomena to compact, overdense gaseous structures in the interstellar medium.  Desh contends that the observations of large differences in the optical depth on two lines of sight separated by a distance $x_0$ are not due to ``clouds'' with physical sizes $x_0$, but rather to the statistics of the optical depth structure function in a medium with a spatial power law in neutral hydrogen density.  Desh emphasized that the optical depth structure function at a spatial lag $x_0$ (essentially what is being measured) is not determined solely by structures with size $x_0$, but in general contains contributions from density structures on all scales.  Desh's presentation drew largely on previous results \cite{Deshpande00} for the TSAS discussion.  He also claimed that a similar reasoning would apply to path integrals through the ionized ISM which produces ESE and other strong scintillation phenomena.  This last contention is not obvious because optical depth is directly describable as a stochastic integral, but scintillation phenomena must be described by wave propagation through a random medium.  

Desh's proposal is controversial because it differs so substantially from the commonly-held view that TSAS, ESE, and related phenomena are due to discrete and extraordinary structures in the interstellar medium.  In contrast, Desh suggests that the observations are essentially statistical fluctuations resulting from a medium with a wide range of spatial scales.  If he is correct, the highly dense, overpressured objects which attracted so much discussion in this meeting would be nonexistent, or at least possess much less extreme properties.  This disturbing prospect should motivate the workshop participants (and others) to explore the mathematics of optical depth differences in a turbulent medium, and scrutinize their data for independent evidence of high pressures and densities in the ISM, or alternatively, indications that such extremes are absent.

\section{Future Research Directions}

One of the useful features of a workshop like this is the guidance it
can provide for future research investigations, the construction of new
instrumentation, and the development of new computer codes.  The
following are the issues which we think should attract the attention of
the community in the next five years or so.

\begin{itemize}

\item As discussed by Dave Meyer, the upgrade to the HST should permit
measurements of column density variations to more, and more
closely-spaced stars, including  binaries and  globular
clusters that have not been observed until now.  This development will give us our best chance to visualize
the features responsible for spatially-variable absorption.  

\item Future developments in radio astronomy would help in improving our
understanding of both the TSAS and ESE phenomena.  The Expanded Very Large Array (EVLA) in the short
term, and the Square Kilometer Array (SKA) in the long term will allow measurements of hydrogen
absorption along more lines of sight to extragalactic radio sources and
pulsars.  

\item The presentations and discussions at this meeting indicated that
simulators should try and quantify the population statistics of the
small-scale density structures in the simulations. They should analyse their simulations so as to produce numbers on filling factors and mass fractions as a function of density threshold.  Extraction of information on clump lifetimes would also be of interest.  All of this information can 
help us decide if highly transient, overpressured ``clouds'' in the simulations  are compatible
with the statistics on lines of sight occupied by TSAS and ESE.   

\item A turbulent plasma is characterized by more than density and its
variations.  The more interesting fluid variables are flow velocity,
magnetic field, vorticity, and the like.  Optical and radio observers
should be encouraged to ``get sophisticated'' and think about ways in
which observations might yield new fluid variables such as vorticity.   

\item A related appeal (which was brought up in the group discussion at
the end of the meeting) is for simulators to develop diagnostics which
would permit a more direct comparison with observations.  One way would
be to calculate path integrals through their simulations, or at a more
advanced level, spectral line profiles.  Such diagnostics would
facilitate the conversation between observers and theorists, and provide
information the observers need.  As Jim Cordes said, speaking of the
relationship between the scattering measure (defined as the path
integral of the turbulence parameter $C_N^2$ (proportional to the
variance of the density fluctuations)) and the emission measure, ``if
you have emission measure, you can turn off the scattering measure, but
you can't have a scattering measure and turn off the emission measure''.

\end{itemize}

\acknowledgements The authors are glad to acknowledge useful comments
from Adriana Gazol. All the conference participants appreciated and will
long remember the excellent organization and arrangements made by NRAO,
as well as the splendid environment of Socorro and central New Mexico.
We particularly want to recognize the fine work of Terry Romero and
Miller Goss. EVS thanks CONACYT for support under grant 47366-F. SRS
thanks the US National Science Foundation for support under grant
ATM03-54782.


\end{document}